\documentclass[aps,preprint,showpacs,floats,epsf,epsfig,nofootinbib,12pt]{revtex4}
\usepackage{graphicx}% Include figure files
\usepackage{epsfig}
\usepackage[dvips]{color}
\usepackage{subfigure}

\def\beq{\begin{eqnarray}}
\def\eeq{\end{eqnarray}}

\def\ra{\rangle}

\begin{document}

\title{Estimating decay rate of $X^{\pm}(5568)\to B_s\pi^{\pm}$ while assuming them to be molecular states}

\vspace{1cm}

\author{ Hong-Wei Ke$^1$\footnote{khw020056@hotmail.com} and
        Xue-Qian Li$^2$\footnote{lixq@nankai.edu.cn}  }

\affiliation{  $^{1}$ School of Science, Tianjin University, Tianjin 300072, China \\
  $^{2}$ School of Physics, Nankai University, Tianjin 300071, China }

\vspace{12cm}

\begin{abstract}
Discovery of $X(5568)$ brings up a tremendous interest because it
is very special, i.e. made of four different flavors.  The D0
collaboration claimed that they observed this resonance through
portal $X(5568)\to B_s\pi$, but unfortunately, later the LHCb,
CMS, CDF and ATLAS collaborations' reports indicate that no such
state was found. Almost on the Eve of 2017, the D0 collaboration
reconfirmed existence of $X(5568)$ via the semileptonic decay of
$B_s$. To further reveal the discrepancy, supposing $X(5568)$ as a
molecular state, we calculate the decay rate of
$X(5568)\rightarrow B_s\pi^+$  in an extended light front model.
Numerically, the theoretically predicted decay width of
$\Gamma(X(5568)\rightarrow B_s\pi^+)$ is $20.28$ MeV which is
consistent with the result of the D0 collaboration
($\Gamma=18.6^{+7.9}_{-6.1}(stat)^{+3.5}_{-3.8}(syst)$ MeV). Since
the resonance is narrow, signals might be drowned in a messy
background. In analog, two open-charm molecular states $DK$ and
$BD$ named as $X_a$ and $X_b$, could be in the same situation. The
rates of $X_a\to D_s\pi^0$ and $X_b\to B_c\pi^0$ are estimated as
about 30 MeV and 20 MeV respectively. We suggest the experimental
collaborations round the world to search for these two modes and
accurate measurements may provide us with valuable information.

\pacs{12.39.Mk, 12.40.-y, 14.40.Lb, 14.40.Nd}

\end{abstract}

\maketitle

\section{Introduction}

Following discovery of numbers of $X,Y,Z$
particles\cite{Abe:2007jn,Choi:2005,Choi:2007wga,Aubert:2005rm,Ablikim:2013emm,
Ablikim:2013wzq,Ablikim:2013mio,Liu:2013dau,Collaboration:2011gj,Ablikim:2014dxl},
whose exotic behaviors cannot be interpreted by the regular $q\bar
q'$ structures and must be attributed to a new type, either
four-quark states or hybrids structures, the discussion on them
becomes a hot topic of the hadron physics. For the four-quark
states, there are several possibilities: molecular state which is
made of two color-singlet mesons; tetraquark which consists of a
color-anti-triplet diquark and a color-triplet anti-diquark, or a
mixture of the previous two. All the possibilities are under
intensive discussions from various angles.

Mostly, the observed exotic $X,Y,Z$ states are composed of hidden
charm or bottom flavors. In 2016 the D0 collaboration declared to
have observed a new resonance $X(5568)$ at the $B_s\pi^{\pm}$
invariant mass spectrum with the mass and width being $(5567.8\pm
2.9^{+0.9}_{-1.9})$ MeV and $(21.9\pm 6.4^{+5.0}_{-2.5})$
MeV\cite{D0:2016mwd}. Since the decay rate of $X(5568)\to B_s
\pi^\pm$ is much larger than that determined by weak interactions,
one can assure that this is a decay caused by strong interaction.
Since for the strong interaction, flavor components do not change
and the final state includes $B_s$ whose quark-component is $(\bar
b s)$ and $\pi^+$  made of $u\bar d$, so in the final state there
are four different flavors which cannot be created from vacuum,
thus one can confirm that $X(5568)$ is a four-quark state which
consists of $\bar b s u\bar d$ ingredients.
Analysis implies $X(5568)$ to be
an exotic state (if it indeed exists), but whether it is a molecule or a
tetraquark would be another open question and need to be answered by precise measurements
combining with careful theoretical studies.
In this work, we investigate its inner
structure via studying its decay behavior.

Unfortunately, the LHCb collaboration\cite{Aaij:2016iev}, the CMS
collaboration of LHC\cite{Sirunyan:2017ofq}, the CDF collaboration
of Fermilab\cite{Aaltonen:2017voc}  and the ATLAS Collaboration of
LHC\cite{Aaboud:2018hgx} claimed that no such decay mode was
detected. Of course, all experimentalists are very careful, so
that they only offered upper bounds on the decay channel. Just on
the Eve of new year, the D0 collaboration declared that $X(5568)$
was re-confirmed in the portal $X(5568)\to B_s\pi^{\pm}$ via a
sequent semileptonic decay of $B_s^0\to
\mu^{\pm}D_s^{\mp}$\cite{Abazov:2017poh} and the result is
consistent with the previous data which were obtained with $B_s\to
J/\psi\phi$, but the measured width is slightly shifted to
$18.6^{+7.9}_{-6.1}(stat)^{+3.5}_{-3.8}(syst)$ MeV. The acute
discrepancy among the experimental groups stimulates a dispute.
Because $X(5568)$ may be the first observed exotic state
possessing four different flavors, studies on it (both theoretical
and experimental) are of obvious significance for getting a better
understanding of the quark model.

In
literature\cite{Guo:2016nhb,Yang:2016sws,Chen:2016ypj,Albaladejo:2016eps,Agaev:2016urs,Stancu:2016sfd,Wang:2016mee,Liu:2016ogz,
Wang:2016tsi,Zhang:2017xwc,Kang:2016zmv,Chen:2016mqt,Xiao:2016mho,Lu:2016kxm,Wang:2018jsr},
there are different opinions which originate from different
considerations. In various models, the spectrum of $X(5568)$ was
computed to be compared with the measured value. Naively, by its
decay width it seems to be a molecular state of
$BK$\cite{Agaev:2016urs,Wang:2018jsr} and its binding energy is
about 205 MeV which is a bit too large for binding two mesons into a hadronic molecule
based on our intuition, thus an alternative suggestion is that it
is a tetraquark\cite{Liu:2016ogz,Wang:2016mee,Stancu:2016sfd}. The
authors of Ref.\cite{Burns:2016gvy} regard that neither a
molecular state nor a tetraquark can explain the data, so they
consider  that $B_s\pi$ is produced in an electroweak decay where
an extra hadron is also created, but evades detection.

In this work, accepting the D0 analysis that $X(5568)$ indeed
exists, we would ask which structure is more preferred by the
nature, it should be answered by fitting more data besides the mass
spectrum, namely one needs to investigate its decay behaviors. Thus a
careful computation on its decay rate is absolutely necessary even
though such a calculation is somehow model-dependent. In fact, a few
groups of authors assumed $X(5568)$ as a tetraquark and computed
the rate of $X(5568)\to B_s\pi$ in terms of the QCD sum
rules\cite{Agaev:2016ijz,Dias:2016dme,Wang:2016wkj}.

%By analyzing its
%production and decay behaviors one can obtain knowledge on
%the inner structure of $X(5568)$.
Different inner structures may result in different decay rates for
a designated channel. Theoretically assigning the molecular
structure to $X(5568)$, we can predict its decay rate to $B_s\pi$.
Since strong interaction is blind to quark flavors, the running
effective coefficients for $b$ and $c$ quarks do not deviate much
from each other. By the heavy flavor symmetry, one believes that
at the leading order, the binding energies for $BK$ and $DK$ are
the same and the symmetry breaking should occur at
($\frac{1}{m_c}-\frac{1}{m_b})$
corrections. As noted, the binding energy for $BD$ might be
different from that of $BK$. Even though the SU(2) symmetry between
$c$ and $s$ quarks is not a good one, the deviation does
not prevent us to make a rough estimate on the binding energy.
We will study their decays while they are supposed to be molecular
states and the results can be a cross check for the mysterious
$X(5568)$.

In order to
explore the decay rates of a molecular state, we extend the light
front quark model (LFQM) which has been successfully applied for calculating decay rates
of regular mesons and baryons
\cite{Jaus,Ji:1992yf,Cheng:2004cc,Cheng:1996if,Cheng:2003sm,Choi:2007se,
Hwang:2006cua,Ke:2007tg,Ke:2009ed,Li:2010bb,Ke:2013zs}.
%To study
%the strong decays of $Z_c(3900)$ and $Z_c(4020)$ which were
%assumed to be molecular states of $D\bar D^*(D^*\bar D)$ and
%$D^*\bar D^*$\cite{Ke:2013gia,Ke:2016owt} whose
%constituents are two mesons instead of a quark and an antiquark, we extended
%the light front quark model. As is well known, the two constituent mesons interact by
%exchanging corresponding mesons\cite{Feng:2011zzb}. By fitting
%relevant processes and some relations, the effective coupling
%constants have been obtained.
Using the method and the parameters  obtained by fitting well measured
data, we deduce the
corresponding transition matrix element and estimate the decay
widths of $X(5568)\rightarrow B_s\pi^+$. Then, we further estimate decay rates of
$X_a\rightarrow D_s\pi^0$ and $X_b\rightarrow B_c\pi^0$ in terms of
the same method where $X_a$ and $X_b$ are the molecular states consisting of $DK$
and $BD$ constituents respectively.

%In the light front framework, we calculate the transition amplitude between
%the initial state and the heavy final state. As $q^+=0$
%condition is set, it requires $q^2<0$. It implies that the light final
%meson is not on-shell, thus we obtained corresponding amplitude in a space-like frame.
%Thus one needs to extrapolate analytically it from the un-physical
%space-like region to the time-like region to gain the physical
%one. With that amplitude we would be able to calculate the decay
%width. The numerical result surely provides us with information
%about the structure of $X(5568)$.

After the introduction we derive the amplitude for transition
$X(5568)\rightarrow B_s\pi^+$, $X_a\rightarrow D_s\pi^0$ and
$X_b\rightarrow B_c\pi^0$  in section II. Then we numerically
evaluate their decay widths in section III. In
the last section we discuss  the numerical results and draw our
conclusion. Some details about the approach are collected in the
Appendix.

\section{the strong decays  $X(5568)$, $X_a$ and $X_b$}
\subsection{the strong decays  $X(5568)\rightarrow B_s\pi^+$}
In this section we calculate the decay rate of
$X(5568)^+\rightarrow B_s\pi^+$, while assuming $X(5568)$ as a
$B\bar K$ molecular state whose quantum number $I(J^P)$ is
$0(0^+)$, in the light-front model. Because of successful
applications of the method to study strong decay processes of
molecular states\cite{Ke:2013gia} we apply the the framework to
the present case. The configuration of the concerned $BK$
molecular state is $\frac{1}{\sqrt{2}}(B^0K^++B^+K^0)$
 \cite{Chen:2016ypj}.
\begin{figure}
        \centering
        \subfigure[~]{
          \includegraphics[width=8cm]{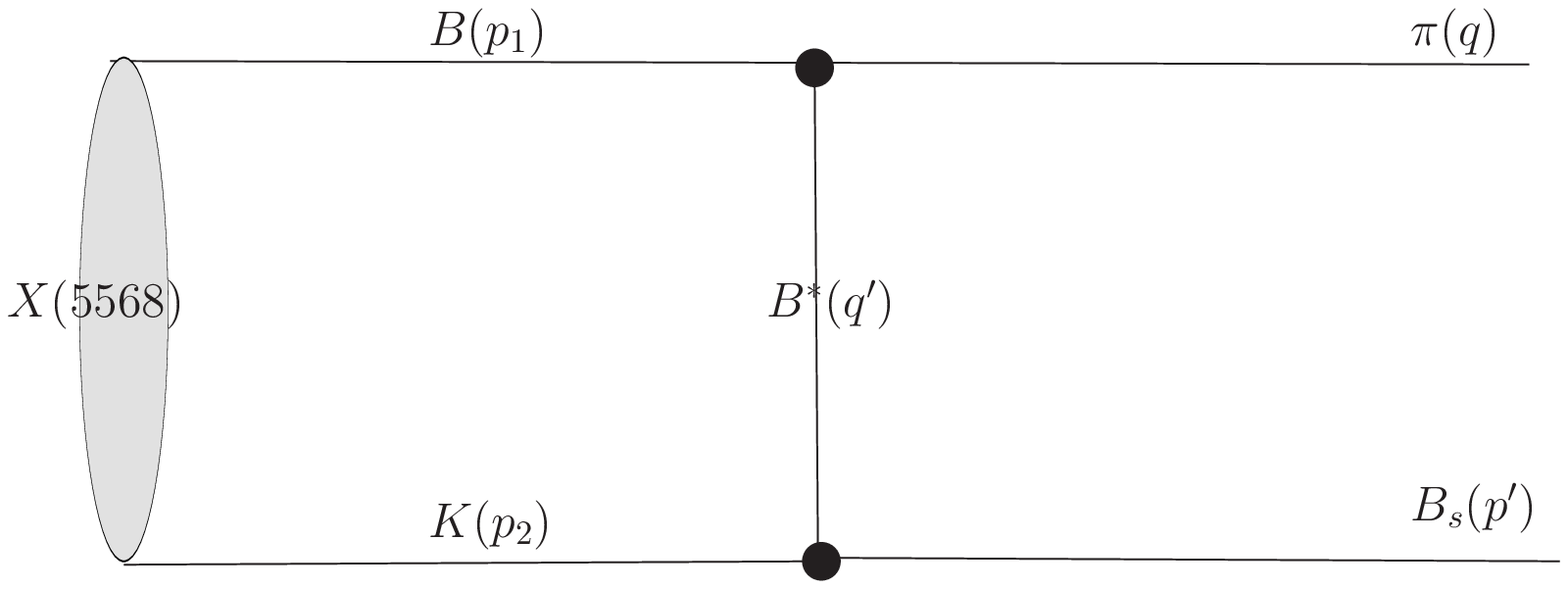}}
        \subfigure[~]{
          \includegraphics[width=8cm]{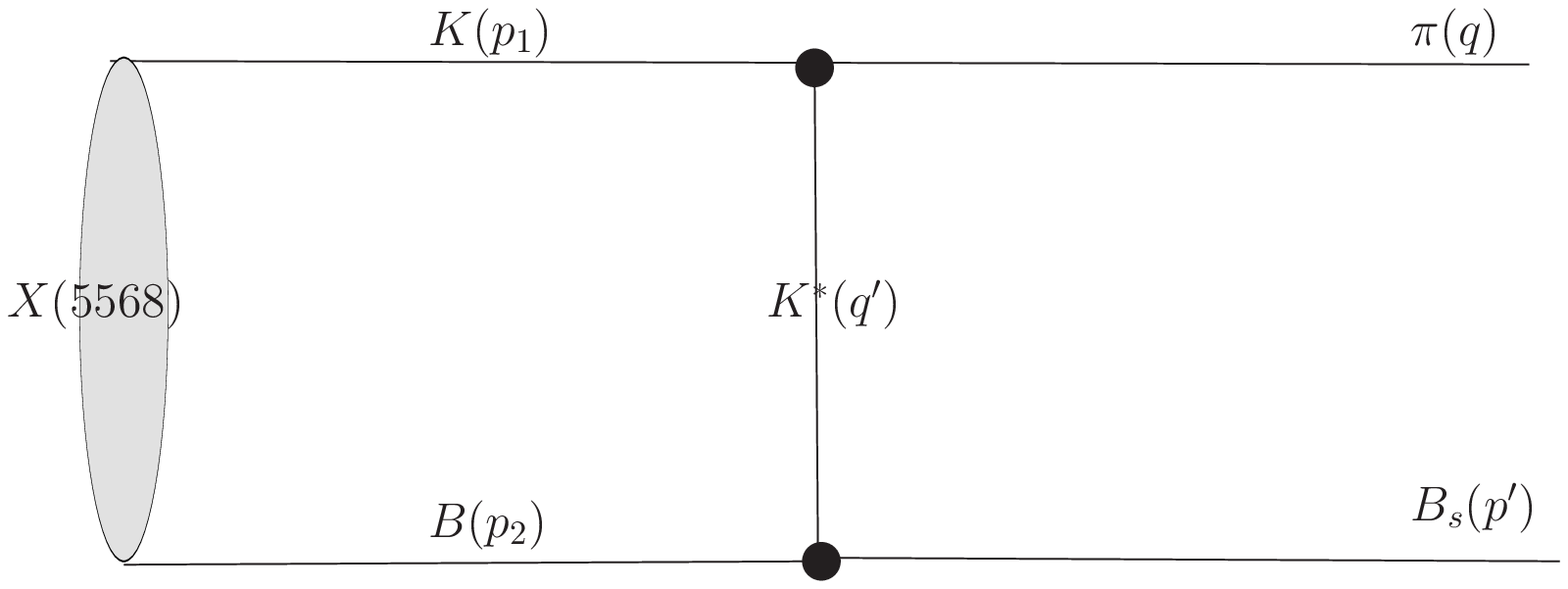}}
\caption{Strong decays of $X(5568)$ .}
        \label{fig1}
    \end{figure}
The Feynman diagrams for $X(5568)$ decaying into $B_s\pi^+$ by
exchanging $B^{*0}\,(\bar B^{*0})$ or $K^{*+}\,(K^{*-})$  mesons
are shown in Fig.\ref{fig1}.

Following Ref.\cite{Cheng:2003sm}, the hadronic matrix
element corresponding to the diagrams  in Fig.\ref{fig1} is written as
\begin{eqnarray}
{\mathcal{A}}_{1}=i\frac{1}{(2\pi)^4}\int d^4
p_1\frac{H_{A}(S^{(a)}+S^{(b)})}{N_1N_1'N_2}
\end{eqnarray}\label{eq1}
with
\begin{eqnarray*} S^{(a)}&&=-i{g_{_{B
B^*\pi}}g_{_{K B^*B_s}}}g_{\alpha\beta}(p_1+q)^\alpha
(2P-p_1-q)^{\beta}
\mathcal{F}(m_1,p_1)\mathcal{F}(m_2,p_2)\mathcal{F}^2(m_{B^*},q'),\\
 S^{(b)}&&=-2i{g_{_{K
K^*\pi}}g_{_{K^* BB_s}}}g_{\alpha\beta}(p_1+q)^\alpha
(2P-p_1-q)^{\beta}
\mathcal{F}(m_1,p_1)\mathcal{F}(m_2,p_2)\mathcal{F}^2(m_{K^*},q'),
\end{eqnarray*}
where $N_1=p_1^2-m_1^2+i\varepsilon$,
 $N_1'={q'}^2-m_{q'}^2+i\varepsilon$,
$N_2=p_2^2-m_2^2+i\varepsilon$ and $P$ stands for the momentum of
$X(5568)$. The form factor
$\mathcal{F}(m_i,k^2)=\frac{(m_i+\Lambda)^2-m_i^2}{(m_i+\Lambda)^2-k^2}$
is introduced to compensate the off-shell effect caused by the
intermediate meson of mass $m_i$ and momentum $k$. The concerned
normalized wavefunction of the decaying meson with the assigned
quantum numbers is included in the vertex function  $H$ which is
invariant in the four-dimensional space-time. In fact, for a practical
computation their exact forms are not necessary, because after
integrating over $dp_1^-$ the integral is reduced into a
three-dimensional one, and then $H$ is replaced by $h$ whose
explicit form is calculable in the light-front frame. In that
frame the momentum $p_i$ is written in terms of its components as
($p_i^-,p_i^+,{p_i}_\perp$) and integrating out $p_{1}^-$ with the
method given in Ref.\cite{Cheng:1996if} one has
\begin{eqnarray}\label{vf9.2}
\int d^4p_1 \frac{H_{A}S}{N_1N_1'N_2}\rightarrow-i\pi\int
dx_1d^2p_\perp\frac{h_{A}\hat S}{x_2 \hat{N_1}\hat{N_1'}} ,
\end{eqnarray}
with
\begin{eqnarray*}
&&\hat{N}_1=x_1({M}^2-{{M}_0}^2),\\
&&\hat{N}_1^{'}=x_2q^2-x_1{{M}_0}^2+x_1M'^2+2p_\perp\cdot
q_\perp,\\&&h_{A}=\sqrt{{x_1x_2}}(M^2-{M}_0^2)h_{A}'
\end{eqnarray*}
where $M$ is the mass of the decaying meson and  $M'$ is the mass
of the heavier one of the two produced  mesons. In the expression,
$q$ is the four-momentum of the lighter meson of the decay
products, while calculating the hadronic transition matrix
element, we deliberately let $q^2$ vary within a reasonable range,
then while obtaining the partial width of
$X(5568)\to B_s\pi$, we set $q^2$ to be the on-shell mass of the
produced pion as $m_{\pi}^2$. The factor
$\sqrt{x_1x_2}(M^2-{M}_0^2)$ in $h_{A}$ was introduced in
literature\cite{Cheng:2003sm}. The explicit expressions of the
effective form factors $h_{A}'$ are presented in the Appendix for
readers' convenience.

Since we calculate the transition in the $q^+=0$ reference frame the zero
mode contributions  which come from the residues of virtual pair
creation processes, were not included. To involve them,
 ${p_1}_\mu$ and ${p_1}_\nu$  in $s^a$ must be
replaced by appropriate expressions as discussed in
Ref.\cite{Cheng:2003sm}, that is
\begin{eqnarray}\label{eq2}
&&{p_1}_\mu\rightarrow\mathcal{P}_\mu A^{(1)}_1+q_\mu A^{(1)}_2
\end{eqnarray}
where $\mathcal{P}=P+P'$ and  $q=P-P'$ with $P$ and $P'$ denoting
the momenta of the concerned mesons in the initial and final
states respectively.

For example, $S^{(a)}$ turns into a replaced form as
\begin{eqnarray} \hat{S}^{(a)}&&=\{-{{m_1}}^2 + \left( 1 + {A^{(1)}_1} + {A^{(2)}_1} \right) \,{{M}}^2
- {{M'}}^2 + 3\,{A^{(1)}_1}\,{{M'}}^2 -
  {A^{(2)}_1}\,{{M'}}^2 - {N1} - {A^{(1)}_1}\,{q^2} - {A^{(2)}_1}\,{q^2}\}
  \nonumber\\&&\frac{-i{g_{_{B
B^*\pi}}g_{_{K B^*B_s}}}}{
     {m_{B^*}}^2}
\mathcal{F}(m_1,p_1)\mathcal{F}(m_2,p_2)\mathcal{F}^2(m_{B^*},q').
\end{eqnarray}

Some notations such as $A_i^{(j)}$ and $M_0'$  can be found in
Ref.\cite{Cheng:2003sm}. With the replacement the amplitude
${\mathcal{A}}$ can be calculated numerically.

\subsection{The decay rate of $X_a\rightarrow D_s\pi^0$}
Now we turn to study the decays of molecules with an open charm. The formulas are
similar to that in the case of open-bottom molecules.

\begin{figure}
        \centering
        \subfigure[~]{
          \includegraphics[width=8cm]{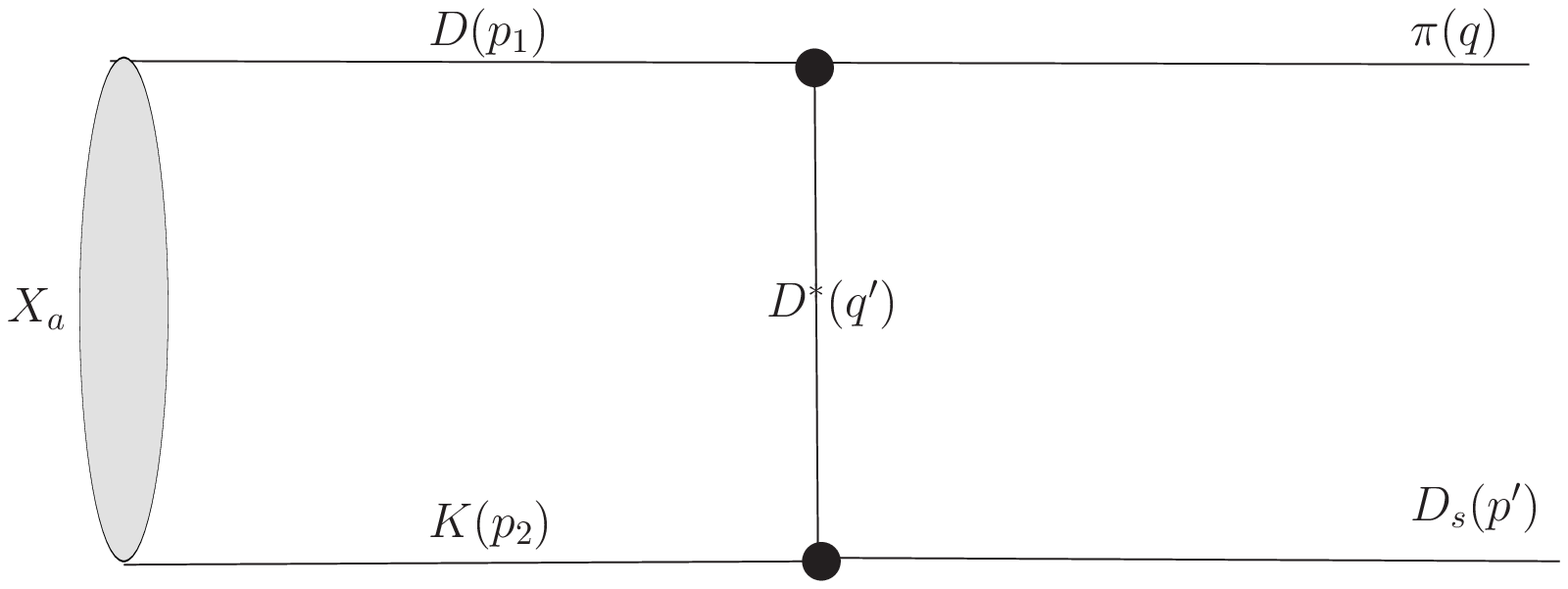}}
        \subfigure[~]{
          \includegraphics[width=8cm]{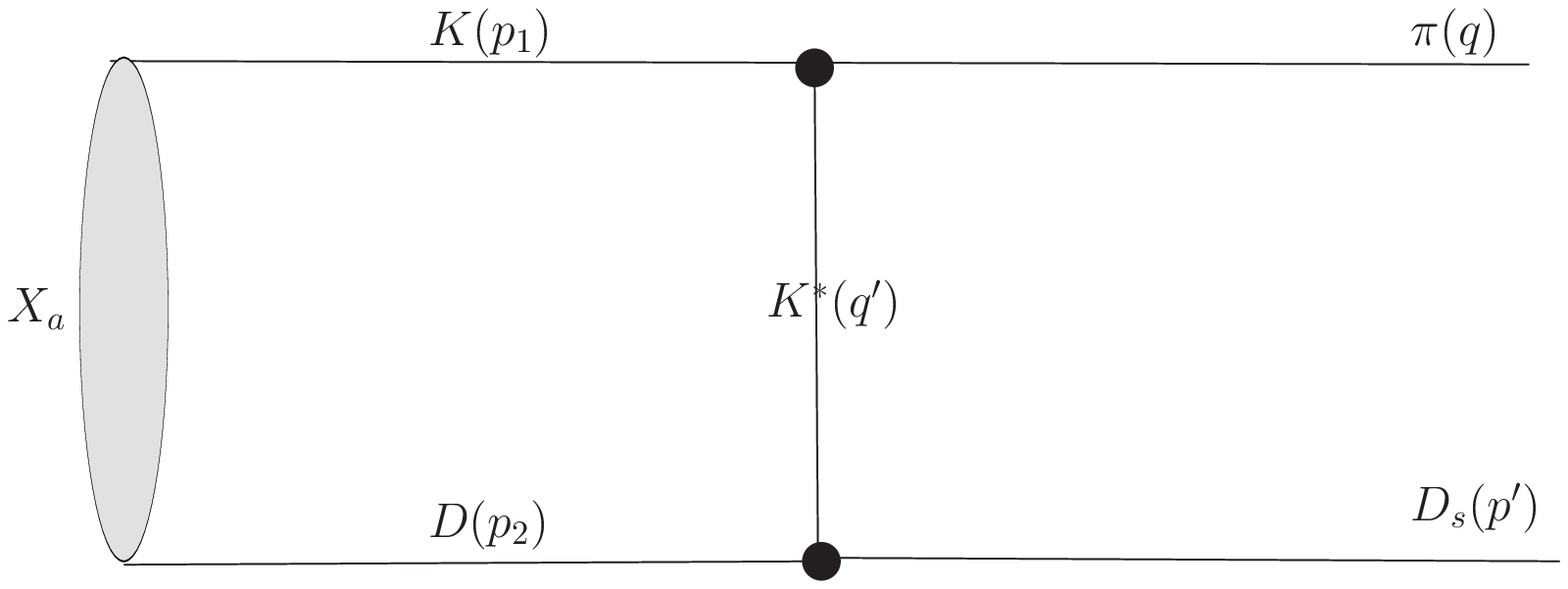}}
\caption{Strong decays of $X_a$.}
        \label{fig2}
    \end{figure}

Due to the quark structure, decay $X_a\rightarrow D_s\pi^0$ realizes
via strong interaction. The supposed molecular state $DK$
($X_a$) is structured as $\frac{1}{\sqrt{2}}(D^0K^++D^+K^0)$. The
Feynman diagrams for $X_a\rightarrow D_s\pi^0$ are shown in Fig.
\ref{fig2}. The corresponding $S^{(a)}$ and $S^{(b)}$ are
\begin{eqnarray*} S^{(a)}&&=-i{g_{_{D
D^*\pi}}g_{_{K D^*D_s}}}g_{\alpha\beta}(p_1+q)^\alpha
(2P-p_1-q)^{\beta}
\mathcal{F}(m_1,p_1)\mathcal{F}(m_2,p_2)\mathcal{F}^2(m_{D^*},q'),\\
 S^{(b)}&&=-2i{g_{_{K
K^*\pi}}g_{_{K^* DD_s}}}g_{\alpha\beta}(p_1+q)^\alpha
(2P-p_1-q)^{\beta}
\mathcal{F}(m_1,p_1)\mathcal{F}(m_2,p_2)\mathcal{F}^2(m_{K^*},q'),
\end{eqnarray*}

\subsection{the decay rate of  $X_b\rightarrow B_c\pi^0$}
\begin{figure}
        \centering
        \subfigure[~]{
          \includegraphics[width=8cm]{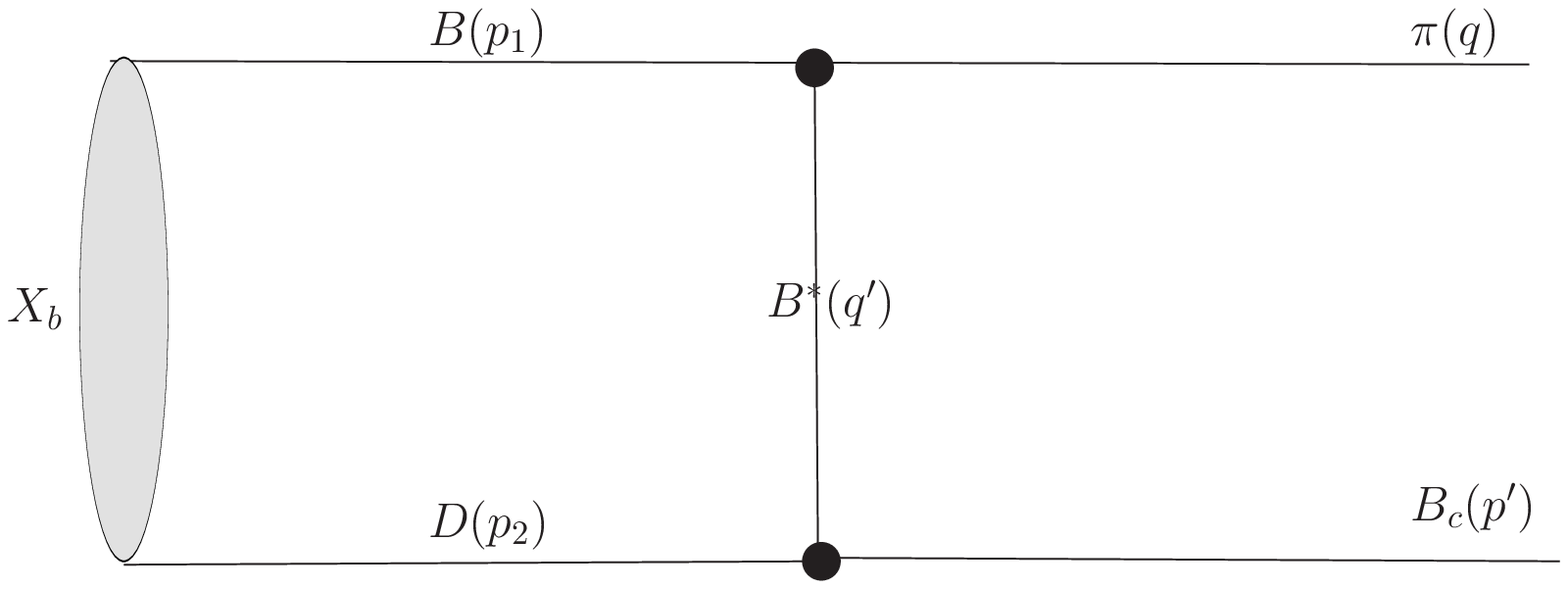}}
        \subfigure[~]{
          \includegraphics[width=8cm]{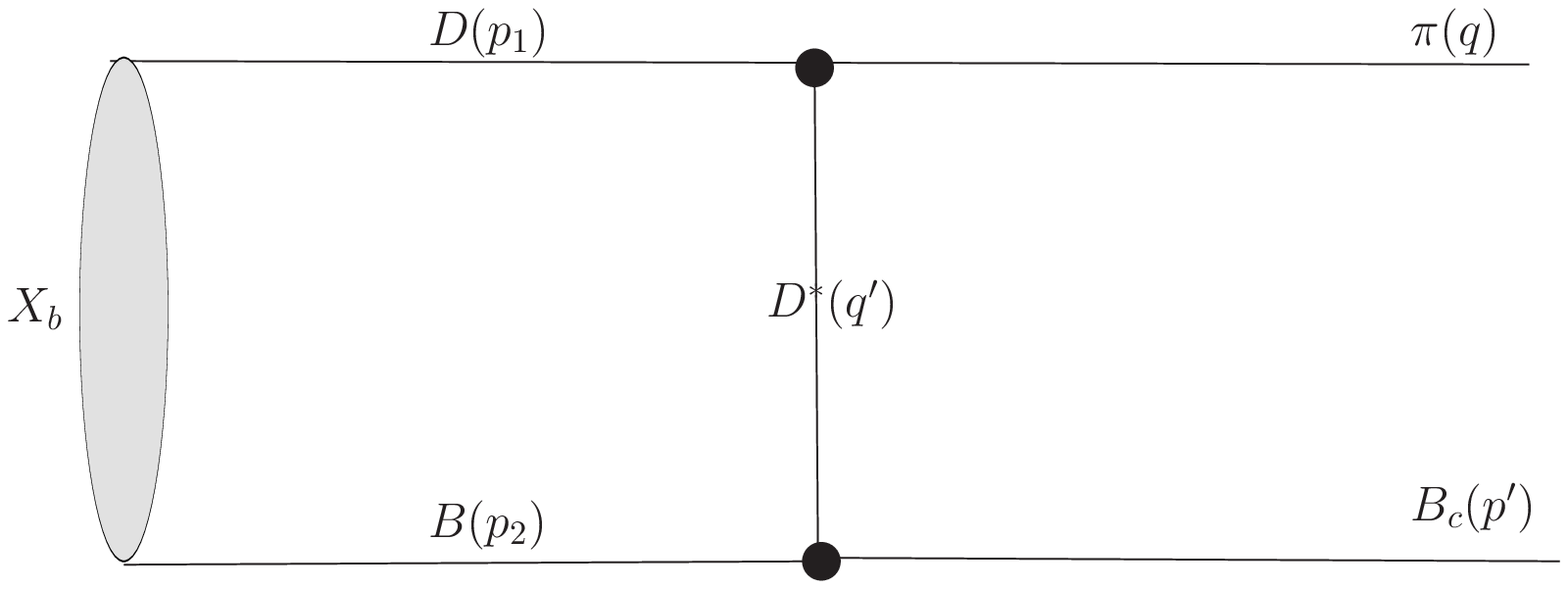}}
\caption{Strong decays of $X_b$.}
        \label{fig3}
    \end{figure}
The molecular state $BD$ ($X_b$) is structured as
$\frac{1}{\sqrt{2}}(\bar D^0B^-+D^-B^0)$. The Feynman diagrams for
the decay $X_b\rightarrow B_c\pi^0$ is shown in Fig. \ref{fig3}.
The corresponding $S^{(a)}$ and $S^{(b)}$ should be modified
as
\begin{eqnarray*} S^{(a)}&&=-i{g_{_{B
B^*\pi}}g_{_{D B^*B_c}}}g_{\alpha\beta}(p_1+q)^\alpha
(2P-p_1-q)^{\beta}
\mathcal{F}(m_1,p_1)\mathcal{F}(m_2,p_2)\mathcal{F}^2(m_{B^*},q'),\\
 S^{(b)}&&=-i{g_{_{D
D^*\pi}}g_{_{D^* BB_c}}}g_{\alpha\beta}(p_1+q)^\alpha
(2P-p_1-q)^{\beta}
\mathcal{F}(m_1,p_1)\mathcal{F}(m_2,p_2)\mathcal{F}^2(m_{D^*},q').
\end{eqnarray*}

\section{numerical results}
\subsection{For $X(5568)\rightarrow B_s\pi^+$ }
In this subsection we present our predictions on the decay rate
of $X(5568)\rightarrow B_s\pi^+$ while all the input
parameters are taken from relevant literatures.

First, we need to calculate the corresponding amplitude which was
deduced in last section. The formula include some parameters which
need to be priori fixed. We use the central value of the observed
resonance peak 5.5678 GeV\cite{D0:2016mwd} as the mass of
$X(5568)$. The masses of the involved mesons are set as
$m_{B}=5.279$ GeV, $m_{B_s}=5.367$ GeV, $m_{\pi}=0.139$ GeV,
$m_{B^*}=5.325$ GeV and $m_{\rho}=0.775$ GeV according to the data
book\cite{PDG12}. The coupling constant $g_{_{K^*K\pi}}$ is
$4.61$\cite{Feng:2011zzb}. About the coupling constants
$g_{_{B^*B\pi }}$, $g_{_{K^*B_sB}}$ and $g_{_{B^*B_sK}}$ one
cannot fix them from the corresponding physical processes at
present but it is natural to conjecture that they would be equal
to $g_{_{D^*D\pi }}$, $g_{_{K^*D_sD}}$ and $g_{_{D^*D_sK}}$
respectively under the heavy quark limit and then they are set as
17.9\cite{Feng:2011zzb}, 3.787\cite{Wang:2006ida} and
2.02\cite{Cheng:2004ru} respectively. The cutoff parameter
$\Lambda$ in the vertex $\mathcal{F}$ was suggested to be 0.88 GeV
to 1.1 GeV \cite{Meng:2007cx}. In our calculation we vary it from
0.88 GeV to 1.1 GeV to study how it affects the numerical results.
$\beta$ in the wavefunction is a free parameter, even though so far
it cannot be precisely determined by phenomenological studies yet, its
value can be roughly estimated to fall within a certain range. We
observe that it should be close to the value for $B$ meson which
was fixed as 0.5329 GeV.
% To see how it affects the final
%theoretical predictions on the decay rates for all the concerned
%modes, we deliberately vary it and plot the dependence of the decay rates of
%$Z_c(3900)\to J/\psi\pi$ and $Z_c(4020)\to J/\psi\pi$ on $\beta$.

Since the amplitude is derived in the reference frame of $q^+=0$ (
$q^2<0$) i.e. in the space-like region, we need to extend it to
the time-like region by means of a normal procedure provided in
literatures. In Ref.\cite{Cheng:2003sm} a three-parameter form
factor  as
\begin{eqnarray}\label{eq23}
\mathcal{A}(q^2)=\frac{\mathcal{A}(0)}{
  \left[1-a\left(\frac{q^2}{M_{X}^2}\right)
  -b\left(\frac{q^2}{M_{X}^2}\right)^2\right]},
 \end{eqnarray}
was employed in order to naturally extrapolate the formula from the space-like region to
the time-like (physical) region.

\begin{table}[!h]
\caption{The   amplitude of $X(5568)\rightarrow
B_s\pi^+$ with three parameters ($\Lambda=0.88$ GeV,).}\label{Tab:t1}
\begin{ruledtabular}
\begin{tabular}{cccc}
   $\beta$ (GeV$^{-1})$    & $A(0)$
& $a$  &  $b$\\\hline0.2&6.64$i$&9.61&15.95\\
0.3&8.71$i$&9.62&15.93\\0.4&9.87$i$&9.41&15.48\\
 0.5329& 10.37$i$&  8.97    &   14.53    \\
 0.6  &  10.36$i$    & 8.72   & 13.97\\
 0.7&10.17$i$&8.32&13.12\\
\end{tabular}
\end{ruledtabular}
\end{table}

%\begin{table}[!h]
%\caption{The decay widths of some modes ($\Lambda=0.88$GeV,
%$\beta=0.2$GeV$^{-1}$).}\label{Tab:t3}
%\begin{ruledtabular}
%\begin{tabular}{cc|cc}
 % decay mode   &  width(MeV)&  decay mode   &  width(MeV)\\\hline
%  $Z_c(3900)\rightarrow J/\psi\pi$  &   0.54 &  $Z_c(4020)\rightarrow J/\psi\pi$  &  1.97   \\
 %  $Z_c(3900)\rightarrow \psi(2S)\pi$  &   2.37&  $Z_c(4030)\rightarrow \psi(2S)\pi$  &   3.67 $\times 10^{-4}$   \\
%   $Z_c(3900)\rightarrow \rho\eta_c$  &  0.045 & $Z_c(4030)\rightarrow \rho\eta_c$
%  &0.030\\
%  $Z_c(3900)\rightarrow DD^*$  &  2.47$\times 10^{-4}$ & $Z_c(4030)\rightarrow DD^*$  &
% 0.14 \\
%  - & - & $Z_c(4030)\rightarrow D^*D^*$  &
% 0.078
%\end{tabular}
%\end{ruledtabular}
%\end{table}

\begin{table}[!h]
\caption{The decay rate of $X(5568)\rightarrow B_s\pi^+$ (
$\beta=0.5329$ GeV).}\label{Tab:t4}
\begin{ruledtabular}
\begin{tabular}{cccccc}
 $\Lambda$(GeV)  &  0.88&0.95&1.0&1.05  &  1.1 \\\hline
 width(MeV) & 20.28&25.40&30.68&36.20 &42.03  \\
\end{tabular}
\end{ruledtabular}
\end{table}

\begin{figure}
\begin{center}
\begin{tabular}{ccc}
\scalebox{0.8}{\includegraphics{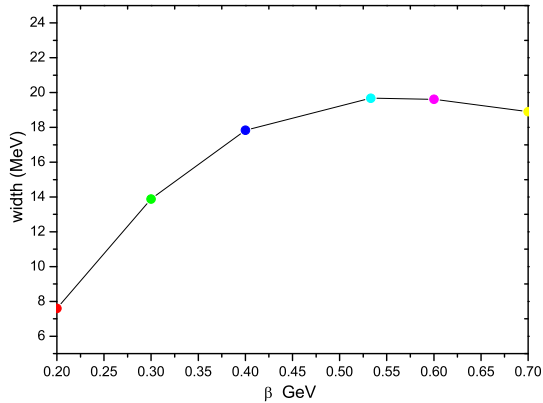}}
\end{tabular}
\end{center}
\caption{the dependence of $\Gamma(X(5568)\rightarrow B_s \pi^+$)
on $\beta$.}\label{Fig6}
\end{figure}

The resultant form factors are listed in table \ref{Tab:t1} and
the dependence of the corresponding decay width
$\Gamma(X(5568)\rightarrow  B_s\pi^+)$ on $\beta$ is illustrated
in Fig. \ref{Fig6}. By the results, we notice that the model
parameter $\beta$ affects the numerical results within a tolerable
range. We also explore the change of the decay width for
different $\Lambda$ values when one sets $\beta=0.5329$ GeV. Since the
channel $X(5568)\rightarrow  B_s\pi^+$ is the dominant portal the
theoretical estimation supports the allegation that $X(5568)$ is a molecular
$BK$ state, especially when $\Lambda=0.88$GeV and $\beta=0.5329$ GeV
the estimated decay width $\Gamma(X(5568)\rightarrow B_s\pi^+)$ is close to the
experimentally measured total width.

\subsection{$X_a\rightarrow D_s\pi^0$ }

Let us turn to discuss the decays of an open-charm molecular state via strong interaction.

%By our general intuition, if $X_a$ is a molecular state
%constructed by $DK$, its binding energy
%should be slightly smaller than
%that for X(5568) (supposing to be a $BK$ molecular state) which is
%205 MeV.

As the reduced mass of the $DK$
system is slightly smaller than that for the $BK$ system, the
corresponding kinematic energy may be larger. Thus, with the same
potential, naively, one would expect a smaller binding energy than
that for the $BK$ system (but not much because the reduced mass is
closer to the mass of the lighter constituent, i.e. the $K$-meson)
where the binding energy is determined to be 205 MeV as it is
considered as an X(5568) molecule. In our concrete numerical
computations, we let the adopted binding vary from 100 MeV to 200
MeV.

%For the transition $X_a\rightarrow D_s\pi^0$ the excganged meson in
%Fig.\ref{fig1}(a) is $D^*$.
The masses $m_D=1.8696$ GeV, $m_{D^*}=2.010$ GeV and
$m_{D_s}=1.968$ GeV are taken from the Databook \cite{PDG12}. A
naive consideration suggests that the parameter $\beta$ is close
to that for $D_s$ which is 0.4395 GeV, meanwhile we set the cutoff
parameter $\Lambda$ to be 0.88 GeV which was obtained in previous
works.   The mass variation covers a range
from 2.164 GeV to 2.264 GeV corresponding to the variation of binding energy
from 100 MeV to 200 MeV.
The results are shown in table
\ref{Tab:t5}.
%As supposing $X(5568)$ to be a $BK$ bound state, one
%can expect that the mass of the $DK$ bound state is around 2.184
%GeV and its width is about 30 MeV.
\begin{table}[!h]
\caption{The decay widths of $X_a\rightarrow D_s\pi^0$ (
$\beta=0.4395$ GeV).}\label{Tab:t5}
\begin{ruledtabular}
\begin{tabular}{ccccccc}
 mass(GeV)  & 2.164&2.184&2.204&2.224 & 2.244&2.264\\\hline
 width(MeV) &21.60&27.47&33.29&39.58&46.27 & 53.43 \\
\end{tabular}
\end{ruledtabular}
\end{table}

\subsection{$X_b\rightarrow B_c\pi^0$ }
Since the $D$ meson is heavier than $K$ meson, assuming the same
arguments on the reduced mass, the binging energy of the bound
state of $BD$ ($X_b$) might be larger than 205 MeV which is the
binding energy of $BK$.
 In our calculation (table \ref{Tab:t6})
we let it vary  from 160 MeV to 240 MeV, which is a typical energy
range (close to $\Lambda_{QCD}$) for binding two mesons into a
compact system. $m_{B_c}=6.2756$ GeV is taken from
Ref.\cite{PDG12} and the parameter $\beta$ adopted for a molecule
with open bottom and charm should be close to $B_c$. Although one
cannot fix it yet from a reliable source at present, we set it to
be a value between 0.631 and 1.257 Gev which are the $\beta$
parameters for $J/\psi$ and $\Upsilon$ respectively, namely we
interpolate the $\beta$ value for $X_b$ to be 0.944 GeV. The
cutoff parameter $\Lambda$ is set as 0.88 GeV. If the mass of $BD$
molecular state is close to 6.929 GeV its width is estimated to be
around 20 MeV.
\begin{table}[!h]
\caption{The decay widths of $X_b\rightarrow B_c\pi^0$ (
$\beta=0.944$ GeV).}\label{Tab:t6}
\begin{ruledtabular}
\begin{tabular}{cccccc}
 mass(GeV)  & 6.909&6.929&6.949&6.969 & 6.989\\\hline
 width(MeV) & 16.51&17.67&18.90&20.20 &21.58  \\
\end{tabular}
\end{ruledtabular}
\end{table}

\section{conclusion and discussions}
Supposing  $X(5568)$ to be a molecular state made by $B$ and $K$
mesons ($BK$), we calculate the decay rate of $X(5568)\rightarrow
B_s\pi^+$  in the light front model. Inside the four-quark
molecule, the two constituents interact by
exchanging corresponding mesons (scalar and/or vector). %We deduce the
%transition amplitude of $X(5568)\rightarrow B_s\pi^+$ in this
%model.
%Since we deduce the amplitude in the $q^+=0$ frame thus it must be
%analytically extended to the time-like region.
%In this process,
%$q^2=m_{\pi}^2$ where $q$ is the momentum of the $\pi$ meson, is
%very small, thus an extrapolation from unphysical region will not affect the value of the
%amplitude $\mathcal{A}(q^2)$ dramatically.
In this phenomenological study, the model parameters $\Lambda$ and
$\beta$ are not fully determined yet at present, so we vary them
within a reasonable range in the numerical computations.
Numerically when $\Lambda=0.88$ GeV and $\beta=0.5329$ GeV are
chosen, we obtain the rate of $X(5568)\to B_s\pi$ as  $20.28$ MeV
which is consistent with the new data measured by the D0
collaboration $\Gamma=18.6^{+7.9}_{-6.1}(stat)^{+3.5}_{-3.8}$ MeV.
The consistency somewhat supports the allegation that $X(5568)$ is
a molecular state composing of $B$ and $K$ mesons.

As long as $X(5568)$ is a molecular state of $BK$ one can expect
two similar states of $DK$ and $BD$ which are named as $X_a$ and
$X_b$ in this work. The
widths of $X_a$ and  $X_b$ are estimated in the same theoretical framework as
roughly 30 MeV and 20 MeV
respectively. The results do not sensitively depend on the choices
of the binding energies. It is worth of putting effort to search
for $X_a\to D_s\pi^0$ and $X_b\to B_c\pi^0$ reactions in sensitive
experimental facilities. It is of obvious theoretical significance, namely
a definite conclusion would help to clarify if such molecular
states are favored by the Nature.

$X(5568)$ is indeed facing an
eccentric situation, namely, the D0 collaboration reconfirmed
their observation of $X(5568)$ at the channel $B_s\pi^{\pm}$
whereas LHCb, CMS, ATLAS and CDF collaborations all gave negative
reports. The sharp discrepancy might be due to a wrong
experimental treatment, but there is still a slim possibility that both
measurements are reasonable because all the measurements with negative
conclusion only gave upper bounds of the rate. Actually, one should make a
theoretical
investigation towards the mysterious exotic hadron, i.e independent
of the experimental data anyway. As a matter of
fact, from the theoretical aspect, there is no rule to forbid
existence of a four-quark state with four different flavors such
as $X(5568)$. Following the lessons we learned from the structures
of $X,Y,Z$ exotic states, it is natural to assume a molecular
state composed of $\bar b, s, u, \bar d$ whose main decay portal
is $B_s\pi$. In this work we used the LFQM to calculate the decay
rate of such a molecule ($X(5568)$) into $B_s\pi$, while another
group\cite{Wang:2018jsr} has also calculated this rate based on
the molecule assumption in terms of the Bethe-Salpeter equation.
Their results are qualitatively consistent with ours and the data
measured by the D0 collaboration. Interesting, some theoretical
groups calculated the decay rate based on the tetraquark
assumption and obtained results of the same order of magnitude.
All these theoretical studies indicate that
$X(5568)$ still may exist, i.e the possibility cannot be simply negated.
However,
the discrepancy between the D0 collaborations with the others persists and must
be taken serious, a reasonable interpretation might be needed.
We believe that this mist would be clarified by the
efforts of both theorists and experimentalists soon.

\section*{Acknowledgement}
This work is supported by the National Natural Science Foundation
of China (NNSFC) under the contract No. 11375128 and 11675082.

\appendix
\section{the vertex function of molecular state}

 The wavefunction of a molecular state with total
spin $J$ and momentum $P$ is\cite{Ke:2013gia}
\begin{eqnarray}\label{eq:lfbaryon}
 |X(P,J,J_z)\rangle&=&\int\{d^3\tilde p_1\}\{d^3\tilde p_2\} \,
  2(2\pi)^3\delta^3(\tilde{P}-\tilde{p_1}-\tilde{p_2}) \nonumber\\
 &&\times\sum_{\lambda_1}\Psi^{SS_z}(\tilde{p}_1,\tilde{p}_2,\lambda_1,\lambda_2)
  \mathcal{F}\left|\right.
  B(p_1,\lambda_1) K(p_2,\lambda_2)\ra.
\end{eqnarray}
For $0^+$ molecular state of $BK$
\begin{eqnarray}
 \Psi^{SS_z}(\tilde{p}_1,\tilde{p}_2,\lambda_1,\lambda_2)&&=
 C_0 \varphi(x,p_{\perp})\equiv h_{C_0}'
\end{eqnarray}

{where $C_{0}$ is the normalization constants which can be fixed
by normalizing the state\cite{Cheng:2003sm}}
 \begin{eqnarray}\label{A6}
\langle X(P',J',J'_z)
|X(P,J,J_z)\rangle=2(2\pi)^3P^+\delta^3(\tilde{P}'-\tilde{P})\delta_{JJ'}\delta_{J_ZJ_{Z'}},
\end{eqnarray}
{and let the normailization
 $
\int\frac{dxd^2p_\perp}{2(2\pi)^3}\varphi'^*_{L',L'_Z}(x,p_\perp)\varphi_{L,L_Z}(x,p_\perp)=\delta_{_{L,L'}}\delta_{_{L_Z,L'_Z}}
$ hold.}

$C_0$ is fixed by calculating Eq. (\ref{A6})
 \begin{eqnarray}\int\frac{dxd^2p_\perp}{2(2\pi)^3}C_0^2\varphi^*(x,p_\perp)\varphi(x,p_\perp)=1,
 \end{eqnarray}
{ then $C_0$=1. It is noted that $P^2=M_0^2$, $p1\cdot P=e_1M_0$
and $p2\cdot P=e_2M_0$ are used as discussed in
Ref.\cite{Cheng:2003sm}. }

and
$\varphi=4(\frac{\pi}{\beta^2})^{3/4}\sqrt{\frac{e_1e_2}{x_1x_2M_0}}{\rm
exp}(\frac{-\mathbf{p}^2}{2\beta^2})$.

{ All other notations can be found in Ref}.\cite{Ke:2007tg}.
\section{the effective vertices}
 The effective vertices  can be found in
\cite{Feng:2011zzb},
\begin{eqnarray}\label{lagrangian_piDD}
 &&\mathcal L_{B^* B\pi}=\frac{g_{_{B^* B\pi}}}{2\sqrt{2}}(iB^{*\mu\dagger}\vec{\tau}\cdot\vec{\pi} \partial_\mu\bar B-iB^{*\mu\dagger}\vec{\tau}\cdot\partial_\mu\vec{\pi} \bar B+h.c.)
 ,\\
 &&\mathcal L_{K^* K\pi}=\frac{g_{_{K^* K\pi}}}{\sqrt{2}}(iK^{*\mu\dagger}\vec{\tau}\cdot\vec{\pi} \partial_\mu\bar K-iK^{*\mu\dagger}\vec{\tau}\cdot\partial_\mu\vec{\pi} \bar K+h.c.)
 ,\\
 &&\mathcal L_{B^* B_sK}={g_{_{B^* B_sK}}}(iB^{*\mu\dagger}K \partial_\mu\bar B_s^0-iB^{*\mu\dagger}\partial_\mu K \bar B_s^0+h.c.),\\
 &&\mathcal L_{K^* B_sB}={g_{_{K^* B_sB}}}(iK^{*\mu\dagger}B\partial_\mu\bar B_s^0-iK^{*\mu\dagger}\partial_\mu B \bar
 B_s^0+h.c.),
 \end{eqnarray}
where $\vec{\tau}$ is usual Pauli matrix. For more details please
refer to Ref.\cite{Feng:2011zzb}.

\end{document}